\documentclass[preprint,prd,aps,showpacs,amsmath,amssymb]{revtex4}
\usepackage{graphicx,color}
\usepackage{tikz}
\usepackage{dcolumn}
\usepackage{bm}
\usepackage[dvips]{psfrag}
\usepackage{psfrag}
\usepackage{epsfig,afterpage}

\begin{document}
\title{Quantum Brownian motion near a point-like reflecting boundary}
\author{V. A. De Lorenci$\mbox{}^{1,2}$}
\email{delorenci@unifei.edu.br}
\author{E. S. Moreira Jr.$\mbox{}^{3}$}
 \email{moreira@unifei.edu.br}
\author{M. M. Silva$\mbox{}^{1,4}$}
 \email{silva.malumaira@gmail.com}

 \affiliation{$\mbox{}^{1}$Instituto de F\'isica e Qu\'imica,
Universidade Federal de Itajub\'a, Itajub\'a, Minas Gerais 37500-903, Brazil}
\affiliation{$\mbox{}^{2}$Institute of Cosmology, Department of Physics and Astronomy,
Tufts University, Medford, Massachusetts 02155, USA }
\affiliation{$\mbox{}^{3}$Instituto de Matem\' atica e Computa\c{c}\~ao,
Universidade Federal de Itajub\'a, Itajub\'a, Minas Gerais 37500-903, Brazil}
\affiliation{$\mbox{}^{4}$Dipartimento di Fisica, Universit\`{a} di
Roma - La Sapienza, Piazzale Aldo Moro 5, I-00185 Rome, Italy}

\date{\today}

\begin{abstract}
The Brownian motion of a test particle interacting with a quantum scalar field in the presence of a perfectly reflecting boundary is studied in $(1+1)$-dimensional flat spacetime. Particularly, the expressions for dispersions in velocity and position of the particle are explicitly derived and their behaviors examined. The results are similar to those corresponding to an electric charge interacting with a quantum electromagnetic field near a reflecting plane boundary, mainly regarding the divergent behavior of the dispersions at the origin (where the boundary is placed), and at the time interval corresponding to a round trip of a light pulse between the particle and the boundary. We close by addressing some effects of allowing the position of the particle to fluctuate. 
\end{abstract}


\maketitle

\section{Introduction}
A small particle interacting with a stochastic field will be driven through a random path. This effect is known as Brownian motion and  has been found in a variety of different physical systems. The classical example is a particle suspended in a liquid (or gas). 
Not long ago a simplified model of an electric charge interacting with a vacuum state showed that a quantum version of Brownian motion may occur.   
In fact, the study of a charged test particle coupled to  electromagnetic vacuum fluctuations due to the presence of a perfectly reflecting plane boundary reveals that the modified vacuum induces a  Brownian motion on the test particle characterized by the following dispersions on its velocity \cite{ford2004} at instant $\tau$,
\begin{eqnarray}
(\Delta v_x)^2 &=& \frac{e^2}{\pi^2m^2}\frac{\tau}{32x^3}\ln\left(\frac{2x+\tau}{2x-\tau}\right)^2,
\label{larryvz}
\\
(\Delta v_y)^2 &=&  \frac{e^2}{\pi^2m^2}\left[\frac{\tau}{64x^3}\ln\left(\frac{2x+\tau}{2x-\tau}\right)^2
-\frac{\tau^2}{8x^2(\tau^2-4x^2)}\right],
\label{larryvx} 
\end{eqnarray} 
and $(\Delta v_z)^2 =(\Delta v_y)^2$. The test particle is taken to have mass $m$,  electric charge $e$, and it is placed at a distance $x$ from the boundary.  We notice that the above vacuum fluctuations can be negative for some values of $x$ and $\tau$,
and that is a consequence of implementing renormalization where the corresponding Minkowski vacuum contributions are subtracted.  (Along the text units are such that $\hbar=c=1$.)

In Eqs.~(\ref{larryvz}) and (\ref{larryvx}), 
the divergences appearing at x=0 are usually associated with the over idealization of the boundary 
condition ~\cite{ford1998} on the field. Another type of divergence occurs when $\tau=2x$, for any value of $x$, and has also been suggested to be related  ~\cite{ford2004} with the use of a perfectly reflecting plane boundary. 
In the regime of $\tau\rightarrow\infty$ the fluctuations in the velocity of the particle in directions parallel to the plane die off while the  fluctuations in the transverse direction reduce to a function that depends only on the position of the particle, namely $\Delta v_x \sim 1/x$ . 
These investigations were extended to the case of two parallel plane boundaries  \cite{hongwei2004}, and to the case where finite temperature effects take place \cite{hongwei2006}. 
A model for ``noncancellation of vacuum fluctuations'' was more recently proposed  \cite{parkinson2011}, using the same setup of an electric charge near a plane boundary.

As has been pointed out in Refs. \cite{ford2004,hongwei2004}, the calculations leading to the results just described were done by assuming that the effects of the vacuum fluctuations begin in a specific instant of time (sudden switch). 
This aspect was considered later \cite{seriu2008} when a smooth switched on/off mechanism was introduced into the model, leading to significant modifications of the quantum Brownian motion originally  described in Ref. \cite{ford2004}.
Furthermore, it has been conjectured that by associating a ``wave packet'' with the particle, 
smearing effects might take place both in space and time directions. Indeed, 
by considering a wave-packet distribution in time direction,
Ref. \cite{seriu2009} has shown that, in the late-time regime, electromagnetic quantum fluctuations yield a transverse velocity dispersion given by $\bigtriangleup v_x \sim 1/\tau$, which is suppressed as $\tau\rightarrow\infty$.
(Note that the smooth switched on/off mechanism \cite{seriu2008}
also leads to this late-time behavior.)
Clearly, in this case, no stationary behavior remains in contrast to the results obtained when the particle is treated ``classically'' as in Ref \cite{ford2004}.
It should be pointed out that in all cases the divergences mentioned earlier survive. 

In this paper we consider the simplified scenario (a toy model) of a scalar charged test particle interacting with a massless quantum scalar field in $(1+1)$-dimensional spacetime, in the presence of a point-like reflecting boundary. 
Brownian motion of the particle coupled to the vacuum fluctuations of the quantum scalar field is examined. 
As occurs in the case of an electric charge interacting with a quantum electromagnetic field near a reflecting plane boundary, the dispersion in the velocity of the particle presents a divergent behavior at the origin (where the boundary is placed), and at the time interval corresponding to a round trip of a light pulse between the particle and the boundary.
We also examine the consequences when the position of the particle is allowed to fluctuate according to a Gaussian distribution.

\section{Vacuum fluctuations and Brownian motion}
\label{sec1}
We restrict ourselves to the case of a test particle of mass $m$ and charge $g$
interacting with a scalar field $\phi(x,t)$  in $(1+1)$ dimensions. 
A perfectly reflecting boundary is assumed to be at $x=0$ where
we impose the Dirichlet boundary condition on the field, $\phi(x=0,t)=0$.  
In the nonrelativistic limit the equation governing the motion of the particle is given by
\begin{eqnarray}
\label{particle}
m\frac{dv}{dt}=-g\frac{\partial}{\partial x}\phi(x,t).
\end{eqnarray}

We assume that the particle interacts with the scalar field but that its contribution to the total field is negligible. This means that the scalar field is solution of the equation $\Box\phi\approx0$, where $\Box$ is just the d'Alembertian in $(1+1)$ Minkowski spacetime written in Cartesian coordinates and therefore dissipation effects due to backreaction are not being considered in our analysis.
It should be remarked that in certain models
dissipative effects become particularly significant on the late-time regime.
Our analysis will be restricted to early-time regime only. 
In fact, the presence of dissipation does not affect the kind of divergent behavior we wish to study. There is still another assumption, namely,
we assume that the particle is at rest at $t=0$ and that its position does not vary 
significantly over time. Thus, its  position $x$ can approximately be considered constant. 

At a given time $\tau$, the velocity of the particle can be obtained by integrating Eq.~(\ref{particle}) as in
\begin{equation}
\label{esc7}
v=-\frac{1}{m}\int_{0}^{\tau}g\frac{\partial}{\partial x}\phi(x,t)dt,
\end{equation}
which states that $\tau$ is the time interval along which  
the particle has been under influence of the scalar field $\phi(x,t)$,
and for this reason $\tau$ is denoted as the measuring time. Notice that in our approach a ``sudden switching'' was assumed at $t=0$. This assumption can be relaxed by introducing a ``smooth switching'' as in Ref. \cite{seriu2008}. 

We are interested in studying the Brownian motion of the test particle in the vacuum state of the quantum scalar field. In this case $\phi(x,t)$ is taken to be an operator acting on a Hilbert space. In order to study the influence of the quantum field over the motion of the particle, we may use the following method. In Eq.~(\ref{esc7}) the force driven by the quantum field [$-g(\partial\phi/\partial x)$] will be considered as a classical stochastic force. Thus, expectation values associated to the velocity and the position of the particle can be calculated by using the moments of the stochastic force, which will be given in terms of the vacuum expectation values of the quantum field \cite{gour1999,jon02}.  

In the case when $\phi(x,t)$ is a pure quantum quantity, it follows that its vacuum expectation value is zero, 
$\langle v \rangle \doteq\langle 0| v |0\rangle =0$, and the mean squared deviation of the particle velocity $(\Delta v)^2$ coincides with the vacuum fluctuation $\langle v^2\rangle$, i.e., $(\Delta v)^2=\langle v^{2}\rangle-\langle v\rangle^{2} =\langle v^{2}\rangle$. 
However, even in the case when  $\phi(x,t)$ has classical contributions, the mean squared deviation would result in a pure quantum quantity. This can be understood in the following way. 
Suppose the scalar field is expressed as a sum of a classical $\phi_{{}_C}(x,t)$ and a quantum $\phi_{{}_Q}(x,t)$ contributions. Thus, the vacuum expectation value of the field is a classical quantity, $\langle\phi(x,t)\rangle= \langle\phi_{{}_C}(x,t)\rangle + \langle\phi_{{}_Q}(x,t)\rangle =  \langle\phi_{{}_C}(x,t)\rangle =  \phi_{{}_C}(x,t)$. However the correlation function is a pure quantum quantity,
$C(x,t;x',t') \doteq \langle \phi(x,t)\phi(x',t') \rangle -\langle \phi(x,t)\rangle \langle\phi(x',t') \rangle
= \langle\phi_{{}_Q}(x,t)\phi_{{}_Q}(x',t')\rangle$, as the classical contribution to the field exactly cancels in the above subtraction. 
Thus, without loss of generality, we will proceed by considering $\phi(x,t)$ as a pure quantum field.  

Using the above considerations, the mean squared deviation of the particle velocity can be obtained as $(\Delta v)^{2}=\langle v(\tau)v(\tau)\rangle$, where
\begin{eqnarray}
\langle v(t_1)v(t_2)\rangle=\frac{g^{2}}{2m^{2}}\left[\frac{\partial}{\partial x}\frac{\partial}{\partial x'}\!\int_{0}^{t_1}\!dt\!\int_{0}^{t_2}\!dt'
G^{(1)}(x,t;x',t')\right]_{x'=x},
\nonumber
\end{eqnarray}
and $G^{(1)}(x,t;x',t') = \big\langle \phi(x,t)\phi(x',t') + \phi(x',t')\phi(x,t)\big\rangle$ is the Hadamard two-point function. 
We can use the relationship between the Hadamard function and the Feynman propagator $G_{F}(x,t;x',t')$ 
to write the above expression in the form,
\begin{eqnarray}
(\Delta v)^{2}=i\frac{g^{2}}{m^{2}}\left[\frac{\partial}{\partial x}\frac{\partial}{\partial x'}
\!\int_{0}^{\tau}\!dt
\int_{0}^{\tau}\!dt'G_{F}(x,t;x',t')\right]_{x'=x}
\label{esc17}
\end{eqnarray}
where the real part of $G_F$ was discarded as it vanishes identically when the coincidence limit is taken. 

The Feymann propagator for a massless scalar field near a perfectly reflecting boundary is solution of $\Box_{(x,t)} G_{F}(x,t;x',t')=-\delta(t-t')\delta(x-x')$. The eigenfunctions of the operator $\Box_{(x,t)}$, which satisfy the Dirichlet boundary condition at $x=0$, are given by
%
$\psi_{\omega,k}(x,t)=(2\pi^2)^{-1/2}\,{\rm e}^{-i \omega t} \sin(k x)$,
%
where $\omega$ and $k$ are real numbers. The corresponding eigenvalues are $\lambda_{\omega,k} = k^{2} - \omega ^{2}$.
Finally, $G_F(x,t;x',t')$ can be obtained by means of \cite{candelas79}
\begin{equation}
G_{F}(x,t;x',t')=-i\int_{0}^{\infty}d\eta \int_{-\infty}^{\infty} dk\int_{-\infty}^{\infty} d\omega {\rm e}^{-i\lambda\eta}\psi_{\omega,k}(x,t)\psi_{\omega,k}^{*}(x',t').
\nonumber
\end{equation}
Performing the above integrals and renormalizing the result with respect to the Minkowski  spacetime, we obtain the renormalized Feynman propagator
\begin{eqnarray}
G^{R}_F(x,t;x',t') = -\frac{i}{4\pi} \ln\left[(x+x')^2-(t-t')^2\right].
\label{feyn}
\end{eqnarray}
In the above result we have discarded an infinite constant term, which corresponds to the usual infrared divergence occurring in two-dimensional spacetime quantum field theory for massless fields \cite{birrel1982}.  It is emphasized that such a divergent term would in any case disappear when the derivatives in Eq.~(\ref{esc17}) were taken over the propagator. 

By putting Eq.~(\ref{feyn}) into Eq.~(\ref{esc17}), operating with the derivatives,  and using the identity 
$\int_{0}^{\tau}dz\int_{0}^{\tau}dyf(|z-y|)=2\int_{0}^{\tau}(\tau-\eta)f(\eta)d\eta$, we obtain
\begin{eqnarray}
(\Delta v)^2 = -\frac{g^2}{4\pi m^2}\ln\left(\frac{4x^2}{\tau^2-4x^2}\right)^2.
\label{18}
\end{eqnarray}
As we can see, there are two divergences in this result which appear also in the electromagnetic case \cite{ford2004}. One at $x=0$, and another when $\tau=2x$ for any value of $x$. The former is the well-known divergence in quantum field theory when the Dirichlet boundary condition is used. The latter is more subtle and corresponds to the travel time a light signal takes to a round trip between the particle and the plane boundary. In the electromagnetic case \cite{ford2004} it was suggested that this divergence also appears because of the assumption of a rigid perfectly reflecting boundary.
As anticipated, we will not be interested in the late-time behavior of $(\Delta v)^2$, for it would be beyond the domain of applicability of the model here considered. This aspect will be better addressed later. 
\begin{figure}[!hbt]
\leavevmode
\centering
\includegraphics[scale = 1.05]{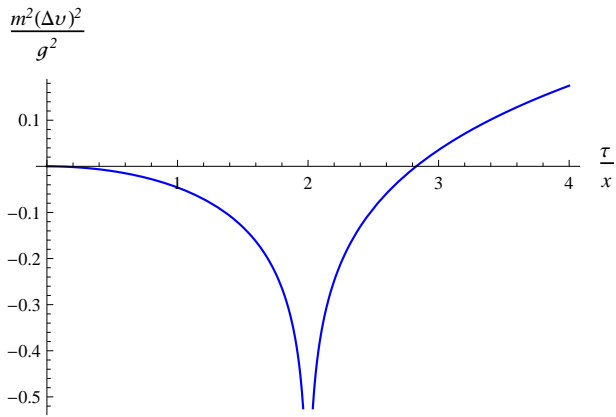}
\caption{{\small\sf (color online). The mean squared deviation of the particle velocity as a function of the measuring time $\tau$ for $x\neq 0$. Besides the divergence at $x=0$, not shown in this figure, this function diverges at $\tau=2x$ as discussed in the text. In the case of a pure quantum scalar field, $(\Delta v)^2 \equiv \langle v^2\rangle$, and subvacuum effects takes place when $0<\tau<2\sqrt{2}x$.}}
\label{eq18tfig}
\end{figure}
The behavior of $(\Delta v)^2$ as a function of $\tau/x$ is depicted in Fig.~\ref{eq18tfig}.
Notice that subvacuum effects ($\langle v^2 \rangle<0$) \cite{hung2012} take place when $\tau^2<8x^2$. In spite of the fact that  
$\langle v^2 \rangle$
is the vacuum expectation value of a positive definite operator,
we should recall  that, formally, $\langle v^2 \rangle = \langle v^2 \rangle_{{}_{Boundary}} - \langle v^2 \rangle_{{}_{Minkowski}}$.
Thus, a negative result is interpreted as a reduction due to the presence of the boundary.  
The origin of the divergence occurring at $x=0$ relies on the boundary condition imposed over the scalar field and it can be understood as follows  \cite{ford1998}. In the renormalization process, the divergences appearing in the propagator derived in the presence of the plane boundary are exactly cancelled out by the corresponding divergences appearing in the Minkowski propagator. However, at $x=0$ we have set $\phi(x=0,t)=0$ (Dirichlet boundary condition). Thus, at this point we are subtracting a finite quantity from an infinity quantity and the renormalization fails.

\begin{figure}[!hbt]
\leavevmode
\centering
\includegraphics[scale = 1.05]{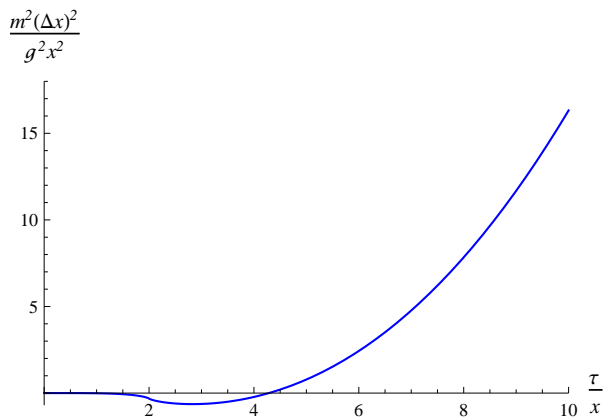}
\caption{{\small\sf (color online). Relative mean squared deviation of the particle position $(\Delta x)^{2}/x^2$ as a function of the measuring time $\tau$ for $x\neq0$. This figure can be used to probe the limit of applicability of the results obtained in this paper. For instance, assuming $m=10g$ we see that, as $\Delta x/x$ must be much smaller than unit, our results are only valid for $\tau\lesssim 10x$.}}
\label{eq47tfig}
\end{figure}

Integration of $v=dx/dt$  allows us to obtain the mean values  $\langle x\rangle=x$, as $\langle v \rangle = 0$,  and $\langle{x}^{2}\rangle = x^{2} + \int_{0}^{\tau}dt_{1}\int_{0}^{\tau}dt_{2} \langle v(t_1) v(t_2) \rangle$.
Thus, calculations similar to those leading to Eq.~(\ref{18}) unveil the following squared mean value $(\Delta x)^{2}= \langle x^ 2 \rangle -\langle x \rangle^2$
of the position of the particle,
\begin{equation}
\nonumber
(\Delta x)^{2} = \frac{g^{2}}{8\pi m^{2}}\left[(\tau^{2}-4x^{2})\ln\left(\frac{\tau^{2}-4x^{2}}{4x^{2}}\right)^{2}-2\tau^{2}\right].
\end{equation}
For an arbitrary position $x\ne 0$ of the particle, $(\Delta x)^{2}$ is a regular function of $\tau$, as depicted in Fig.~\ref{eq47tfig}. Particularly, at $\tau=2x$ we obtain $(\Delta x)^{2}=-({g^{2}}/{\pi m^{2}})x^2$, which is a regular function of $x$.
We notice that this regular behavior also occurs in the case of an electric charge near a reflecting plane boundary, but only in the direction perpendicular to the plane. 

In deriving the above results we have assumed that the particle does not change much its position over time. This assumption can be stated as $\left|(\Delta x)^2/x^2\right| \ll 1$, which implies in the condition
\begin{equation}
\frac{g^{2}}{8\pi m^{2}}\left|\left(\frac{\tau^{2}}{x^{2}}-4\right)\ln\left(\frac{\tau^{2}-4x^{2}}{4x^{2}}\right)^{2}-2\frac{\tau^{2}}{x^{2}}\right|<<1.
\label{49}
\end{equation}
Particularly, at $\tau=2x$, 
\begin{equation}
\label{pos15d}
\frac{g^{2}}{\pi m^{2}}<<1,
\end{equation}
which should hold for any values of $x$ e $\tau$.

We stress that the inequality stated by Eq.~(\ref{49}) imposes a natural limit of validity over our results. For instance, setting $g/m=0.1$,  which obeys Eq.~(\ref{pos15d}), we obtain that $\tau\approx 10x$ appears as an upper limit of validity for the obtained dispersions, as in this case $|(\Delta x)^{2}|/x^{2}\approx 0.16$. 
Taking $g/m=0.01$ we obtain that when $\tau\approx50x$ it results $|(\Delta x)^{2}|/x^{2}\approx 0.11$.
To summarize, the smaller $g/m$ the greater the range of  validity of our results. This conclusion can be understood by analyzing the behavior of the relative dispersion $(\Delta x/x)^2$ in Fig.~\ref{eq47tfig}.

\section{Final remarks}
\label{remarks}
In this work the scalar field was considered as a quantum field while particle and boundary were treated at fixed positions.  A more realistic description of the system should consider the quantum nature of all its components.
We could simulate this aspect by allowing the position of the particle to fluctuate around $x$. For instance, let  $x\rightarrow x + \epsilon$, where the parameter $\epsilon$ is the random variable in the Gaussian distribution $f(\epsilon) = (1/\sqrt{2\pi}\sigma){\rm exp}(-\epsilon^2/2\sigma^2)$ with width $\sigma$. Hence the mean value over $\epsilon$ of velocity dispersion $(\Delta v)^{2}$ can be obtained by means of
$\overline{(\Delta v)^{2}} = \int_{-\infty}^{\infty} (\Delta v)^{2}\,f(\epsilon) d\epsilon.$
\begin{figure}[!htb]
\leavevmode
\centering
\includegraphics[scale = 0.95]{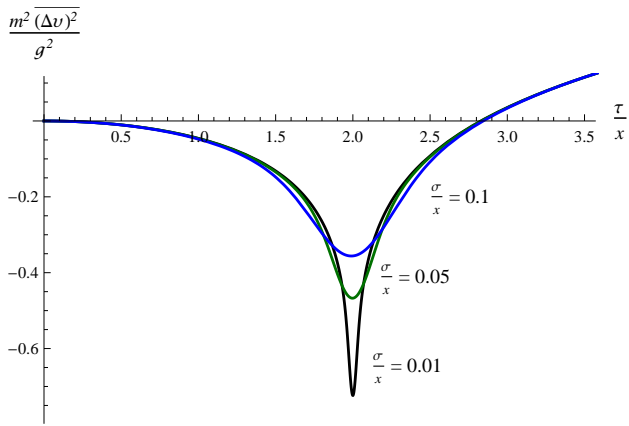}
\caption{{\small\sf (color online).
$\overline{(\Delta v)^{2}}$ as a function of $\tau/x$ for different values of the width $\sigma$. Notice that as $\sigma/x\rightarrow 0$, $\overline{(\Delta v)^{2}}$ tends to the singular $(\Delta v)^{2}$ at $\tau=2x$.}}
\label{fig38t}
\end{figure}
Straightforward calculations [using Eq.~(\ref{18})] show that the above integration can be solved in terms of the generalized hypergeometric function ${_{{}_{2}}F_{{}_2}}\left(1,1;\frac{3}{2},2;z\right)$ \cite{lebedev1972}  and results in a regular function of $x$ and $\tau$, as depicted in Fig.~\ref{fig38t}. 
Notice that the smaller $\sigma$ the more $\overline{(\Delta v)^{2}}$ approaches the case without position fluctuation depicted in Fig.~\ref{eq18tfig}. Particularly, for a given distance $x$ from the boundary,  we obtain
\begin{equation}
\overline{(\Delta v)^{2}} \approx
\frac{g^{2}}{4\pi m^2}\ln\frac{2\sigma^2}{x^2}, \quad \{\tau=2x, \;\; \sigma\rightarrow 0\},
\nonumber
\end{equation}
showing that the depth of the well in Fig.~\ref{fig38t} sharpens as  $\ln\sigma/x$ when $\sigma\rightarrow 0$.

It is to be seen if such a ``position fluctuation'' is enough to regularize velocity dispersions in all directions when higher dimensional models are considered, and if indeed it corresponds to genuine quantum motion of the test particle. These are points that require further analysis.

\acknowledgments
V.A.D.L thanks L. H. Ford for valuable discussions.  This work was partially supported by the Brazilian research agencies
CNPq, FAPEMIG, and CAPES under Scholarship No. BEX 18011/12-8. M.M.S. thanks CAPES for supporting her M.Sc. studies.

\end{document}